\documentclass[prb,preprint]{revtex4-1} 

\usepackage{amsmath}

\usepackage{graphicx}
\usepackage[section]{placeins}
\usepackage{soul}

\usepackage{hyperref}   
\hypersetup{
	colorlinks=true, % 设置超链接颜色
	linkcolor=blue, % 设置链接颜色
	citecolor=blue, % 设置引用颜色
	urlcolor=magenta, % 设置 URL 颜色
	pdfborder={0 0 0} % 设置 PDF 边框（这里设置为无边框）
}

\begin{document}

\title{Revisiting Koehler's experiment of measuring the ratio of the specific heats of air by self-sustained oscillations}

\author{Yujun Shi}
\email{yujunshi@sxu.edu.cn}
\affiliation{School of Physics and Electronics Engineering, Shanxi University, Taiyuan 030006, People's Republic of China}

\author{Xiaoting Feng}
\affiliation{School of Physics and Electronics Engineering, Shanxi University, Taiyuan 030006, People's Republic of China}

\begin{abstract}
We revisit Koehler's experiment, a clever modification of Rüchardt's experiment designed to measure the ratio of specific heats of gas. However, the lengthy and dense analysis shared by Koehler in his 1950 paper may pose challenges to readers due to the complexity of the calculations. Following Koehler's approximation for pressure changes, we explicitly present the model equations as piecewise linear differential systems and qualitatively analyze the periodic solutions from a geometric perspective. This concise and transparent approach addresses a fundamental question about Koehler's experiment: why is the oscillation frequency nearly equal to the Rüchardt frequency? Our analysis avoids intricate calculations and should help educators introduce Koehler’s experiment in general physics laboratory classes. 
\end{abstract}

\maketitle
\section{Introduction}
\begin{figure}[h!]
\includegraphics[width = 0.6\textwidth]{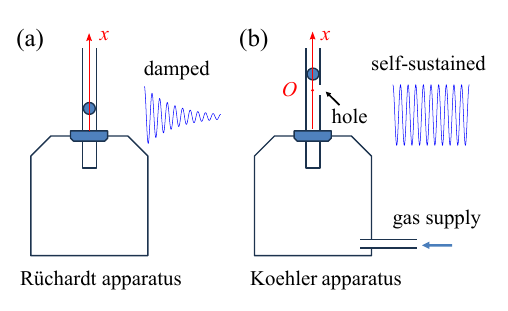}
\caption{(a) and (b) Schematic diagrams of Rüchardt apparatus and Koehler apparatus.}
\label{apparatus}
\end{figure}

Koehler's experiment for measuring the ratio of specific heats $\gamma = C_{p}/C_{v}$ of air (or other gases) is a modification of Rüchardt's experiment, a classic experiment widely used in general physics laboratory courses \cite{Koehler1951, Lange2000}.  As shown in Fig. \ref{apparatus}(a), the apparatus for Rüchardt's experiment consists of a large sealed vessel connected to a vertical tube, in which a closely fitting steel ball can move freely. When the ball is stationary at its equilibrium position, the pressure in the vessel satisfies the equation
\begin{equation}\label{P0}
	 P_0=P_{\mathrm{atm}}+\frac{mg}{\pi R^2}, 
\end{equation}
where $P_{\mathrm{atm}}$ is the atmospheric pressure, and $m$ and $R$ are the mass and radius of the ball, respectively. When the ball is displaced vertically from equilibrium and released, it undergoes rapid oscillatory motion along the $x$ axis (see Fig.~1(a)), inducing pressure fluctuations through the adiabatic compression and expansion of the gas in the vessel. The equation of motion for the ball can be written as
\begin{equation}
	m\ddot{x} = \pi R^2 \,\Delta P.
	\label{eq1}
\end{equation}
In this equation, $x$ is the displacement of the ball from its equilibrium position (positive upward), and $\Delta P$ is the pressure change in the vessel. Assuming an adiabatic process, the gas satisfies
\begin{equation*}
	PV^\gamma = \mathrm{const.}~.
\end{equation*}
Because the volume change induced by the motion of the ball is small compared with the total volume, the adiabatic relation can be linearized about equilibrium, yielding
\begin{equation}\label{eq2}
	\Delta P=-\frac{\gamma P_{0}\Delta V}{V_0}=-\frac{\gamma P_{0}\cdot\pi R^{2}x}{V_0},
\end{equation}
where $V_0$ is the total gas volume when the ball is at equilibrium, including the vessel and a portion of the tube. In practice, $V_0$ can be approximated by the volume of the vessel, since the contribution from the tube is negligible. In this limit, the ball executes simple harmonic motion. The relationship between the period of oscillation $T$ and $\gamma$ is given by
\begin{equation}
	\gamma=\frac{4mV_0}{T^{2}P_{0}R^{4}}.
	\label{gamma}
\end{equation}

Although Rüchardt's  method has the advantages of simplicity and mechanical intuitiveness, it is not as simple as it first appears. Due to frictional damping and gas leakage, the ball executes only a few oscillations before coming to rest, preventing precise measurement of the period.  In 1951, Koehler described an interesting modification of this method, involving a gas-feeding pump and a small hole in the tube (as shown in Fig. \ref{apparatus}(b))\cite{Koehler1951}. The ball is lifted by gas overpressure and, once its vertical position is above that of the hole, falls back as gas leaks. When the gas flow is adjusted to be within a certain range, the ball can oscillate symmetrically about the hole indefinitely. The oscillations closely approximate harmonic motion, with a period still given by Eq.(\ref{gamma}). The advantages of Koehler's method are evident: it allows prolonged measurement operation, thereby improving the ease and precision of the measurements. Therefore, compared to Rüchardt's experiment, Koehler's experiment offers greater advantages for teaching purposes, although it appears to be less widely spread. To encourage educators to explore this method in their course, the detailed parameters of the experimental apparatus employed in this study are provided in Appendix \ref{appendixA}.

In courses using Koehler's experiment, the theoretical introduction typically follows the framework of the original Rüchardt experiment. Students are generally informed that the ball's motion approximates harmonic behavior, as described by Eq.(\ref{gamma}). However, the derivation for Eq.(\ref{gamma}) cannot be identically replicated, since the pressure fluctuations are not only the result of the adiabatic changes in the gas volume, but are also due to the incoming gas and to the gas leakage through the hole. Koehler presented a theoretical explanation for the ball's self-sustained oscillations in another paper \cite{Koehler1950}. Assuming the ball's motion to be sinusoidal, the author used nonlinear dynamical equations to demonstrate that the pressure fluctuations are also approximately sinusoidal, with a frequency closely matching the Rüchardt frequency. However, this analysis is indirect and lengthy, and might be challenging to readers. To our knowledge, no other studies provide a more accessible theoretical analysis of this issue. 

The following sections of this paper will revisit this issue and provide a concise theoretical interpretation of the self-sustained oscillations. The paper is organized as follows. In section~II, following Koehler's approximation for pressure changes, we present the dynamical equations governing the ball's motion. In section~III, we nondimensionalize
the model equations and express them as a three-dimensional piecewise linear differential system. The behavior near the fixed points is then analyzed in a two-dimensional projection. In section~ IV, we qualitatively analyze the properties of the system's periodic solutions from a geometric perspective, demonstrating that the oscillation frequency in Koehler's experiment is nearly equal to the Rüchardt frequency.

\section{Dynamical equations of self-sustained oscillations}
Let $x$ and $v$ denote the vertical coordinate and velocity of the ball, respectively. Let $P$ denote the pressure in the vessel, which is a function of the ball's coordinate $x$ and time $t$, and is assumed to be uniform in the vessel. Let $P_0$ denote the equilibrium pressure, given by Eq.~(\ref{P0}). The dynamical equations are
\begin{subequations} \label{ini_dynamics}
    \begin{align}
&\frac{dx}{dt} =v, \\    
&\frac{dv}{dt} =\frac{(P-P_{0})\pi R^{2}}{m}, \\
&\frac{dP}{dt} =\frac{\partial P}{\partial x}\frac{dx}{dt}+\frac{\partial P}{\partial t} . \label{dP/dt}
\end{align}
\end{subequations}

Eq.(\ref{dP/dt}) shows that the rate of pressure change consists of two parts: the fist term arising from the adiabatic change of the gas volume due to the ball's motion, and the term $\partial P/\partial t$ due to the gas supply and leakage. For the adiabatic process of an ideal gas, 
\begin{equation}\label{dP/dx}
    \frac{\partial P}{\partial x}=\frac{\partial P}{\partial V}\frac{\partial V}{\partial x}=-\frac{\gamma P\pi R^2}V\sim-\frac{\gamma P_0\pi R^2}{V_0}.
\end{equation}
Here also, we have assumed that the ball's displacement induces a negligible change in the vessel's volume: $P/V\sim P_0/V_0$. For $\partial P/\partial t$, we use the approximation from Koehler's paper\cite{Koehler1950}. 

The contribution to $\partial P/\partial t$ due to the gas supply (an aquarium air pump) is approximated as a positive constant
\begin{equation}\label{dP/dt1}
	\left(\frac{\partial P}{\partial t}\right)_{\mathrm{supply}} = jP_{\mathrm{atm}}.
\end{equation}
This approximation is reasonable because the air pump used here operates approximately as a constant-flow source over the relevant pressure range. Assuming a constant volumetric flow rate $Q$, the ideal gas equation
\[
PV=nRT
\]
gives
\[
\frac{dP}{dt}
=
\frac{RT}{V}\frac{dn}{dt},
\]
Since the molar inflow rate satisfies
\[
\frac{dn}{dt}
=
\frac{P_{\mathrm{atm}}Q}{RT},
\]
one obtains
\[
\frac{dP}{dt}
=
\frac{P_{\mathrm{atm}}Q}{V}.
\]
Therefore,
\[
j=\frac{Q}{V}.
\]

For the pump used in the experiment described in the Appendices, the maximum flow rate is approximately
\[
Q\sim7~\mathrm{L/min}
\sim1.2\times10^{-4}~\mathrm{m^3\,s^{-1}}.
\]
For a typical container volume $V=2.5~\mathrm{L}$, the corresponding order of magnitude of $j$ is estimated to be
\[
j\sim 0.05~\mathrm{s^{-1}}.
\]
Moreover, the characteristic timescale associated with the pressure increase, $1/j\sim20~\mathrm{s}$, is typically much longer than the oscillation period of the system, which is of order one second. Hence, over several oscillation cycles, the pressure increase produced by the pump may be regarded as approximately linear in time, which justifies the approximation used in Eq.~(\ref{dP/dt1}).

Similarly, since the pressure variation induced by the gas leakage is also small, the contribution to  $\partial P/\partial t$  due to leakage is approximated as being proportional to the pressure difference between inside and outside the container:
\begin{equation}\label{dP/dt2}
    \left(\frac{\partial P}{\partial t}\right)_{\mathrm{leakage}}=-g(x)(P-P_{\mathrm{atm}}), \quad g(x)=\begin{cases}e+f,&\quad x\geq 0\\e,&\quad x<0\end{cases},
\end{equation}
where the origin of the $x$-axis is defined at the center of the small hole on the tube. When $x<0$, the gas escapes solely through the unavoidable annular gap between the ball and the tube wall. When $x\geq 0$, the gas escapes through both the annular gap around the ball and the small hole. The finite size of the small hole is neglected, resulting in a discontinuity of $g (x)$ at $0$. However, this effect is minor and not crucial to the main point. By substituting Eqs.~(\ref{dP/dx}), (\ref{dP/dt1}), and (\ref{dP/dt2}) into Eqs.~(\ref{ini_dynamics}), the final dynamical equations are derived as

\begin{subequations} \label{final_dynamics}
    \begin{align}
&\frac{dx}{dt}=v, \\
&\frac{dv}{dt}=\frac{(P-P_{0})\pi R^{2}}{m}, \\
&\frac{dP}{dt}=-\frac{\gamma P_{0}\pi R^{2}}{V_0}v+jP_{\mathrm{atm}}-g(x)(P-P_{\mathrm{atm}}).
\end{align}
\end{subequations}
When the term $jP_{\mathrm{atm}}-g(x)(P-P_{\mathrm{atm}})$ is removed, the equations reduce to those describing the harmonic motion in the ideal Rüchardt experiment. The Rüchardt frequency is given by $\omega^2=\pi^2\gamma P_0R^4/(mV_0) $(equivalent to Eq. (\ref{gamma})). The term involving the piecewise function $g(x)$ introduces nonlinearity. As a result, the equations can no longer be solved analytically.

\section{Preliminary qualitative analysis}

We first nondimensionalize Eqs. (\ref{final_dynamics}) as follows:
\begin{subequations} \label{dimensionless_dynamics}
    \begin{align}
&\frac{dy}{d\tau} =u, \\
&\frac{du}{d\tau} =Ap, \\
&\frac{dp}{d\tau} =-\frac1Au-G(y)p+ [J-G(y)(p_0-1)] , 
\end{align}
\end{subequations}
where $\tau=\omega t,y=x/R, u=v/(R\omega), p=(P-P_0)/P_{\mathrm{atm}}, p_0=P_0/P_{\mathrm{atm}}, A=(\pi RP_{\mathrm{atm}})/(m\omega^{2}), J=j/\omega$, and  
\[G(y)=\frac{g(x)}{\omega}=\begin{cases}G_+ = (e+f)/\omega,\quad&y\geq 0,\\G_- = e/\omega,\quad&y<0.\end{cases}\]

Equation (\ref{dimensionless_dynamics}) is a piecewise linear differential system (PWLD systems). This system consists of two linear subsystems separated by the plane defined by $\{y = 0\}$, referred to as the separation plane. They can be written in matrix form as
\begin{equation} \label{matrix form 3D_equation}
\dot{\mathbf{z}}=\begin{cases}M_+\mathbf{z}+\mathbf{e}_+&\mathrm{if}\quad y\geq0 \\M_-\mathbf{z}+\mathbf{e}_-&\mathrm{if}\quad y<0\end{cases}
\end{equation}
with 
\[\mathbf{z}=\begin{pmatrix}y\\u\\p\end{pmatrix},\quad
M_+=\begin{pmatrix}0&1&0\\0&0&A\\0&-\frac1A&-G_+\end{pmatrix},\quad M_-=\begin{pmatrix}0&1&0\\0&0&A\\0&-\frac1A&-G_-\end{pmatrix}\] and
\[ \mathbf{e}_+=\begin{pmatrix}0\\0\\J-G_+(p_0-1)\end{pmatrix},\quad
\mathbf{e}_-=\begin{pmatrix}0\\0\\J-G_-(p_0-1)\end{pmatrix}.\]
PWLD systems\cite{Llibre2018} are widely used to model real-world phenomena in fields such as electrical circuits, control systems, impact and friction mechanics, and biological systems. However, analyzing periodic solutions, which often correspond to steady-state responses in general PWLD systems, is notoriously challenging \cite{Jeffrey2018}. This difficulty arises because PWLD systems switch their behavior at the separation plane, leading to discontinuities and distinct sets of governing equations in different regions of the phase space. Accurately handling transitions across these boundaries often requires solving implicit or transcendental equations, which can be particularly challenging or even infeasible in practice. Despite these challenges, the following sections provide an approximate analysis showing that, if periodic solutions exist—i.e., solutions for which the ball passes through $x=0$ at regular time intervals—the oscillation frequency is very close to the Rüchardt frequency.

Analyzing fixed points (equilibrium points) yields valuable insights for dynamical systems. In both half-spaces $\{y\geq 0\}$ and $\{y< 0\}$, Eqs.(\ref{dimensionless_dynamics}) do not exhibit global fixed points. However, note that on the right-hand side of the equations, the variable $y$ appears only within $G(y)$, whose constant value is piecewise determined by $y$. Thus, $y$ can be treated as a parameter, allowing for an analysis of the system in the $(u, p)$ projection plane for each half-space, $\{y\geq 0\}$ and $\{y< 0\}$:
\begin{equation} \label{u-p subsystem}
    \begin{aligned}
&\frac{du}{d\tau}=Ap,\\
&\frac{dp}{d\tau} =-\frac1Au-Gp+[J-G(p_0-1)] ,
\end{aligned}
\end{equation}
where $G \equiv G(y)$. Setting $du/d\tau = 0$ and $dp/d\tau = 0$, we obtain the fixed point of this subsystem, denoted by $(u^*, p^*)$, as
\begin{equation}\label{fixed_point}
u^* = A\left[J - G(p_0 - 1)\right], \quad p^* = 0.
\end{equation}
To analyze the local behavior near the fixed point, we introduce small perturbations
\[
u = u^* + \eta(\tau), \quad p = p^* + \xi(\tau),
\]
where $\eta(\tau)$ and $\xi(\tau)$ are small deviations. Substituting into Eqs.~(\ref{u-p subsystem}), we obtain
\begin{equation}\label{Jacobian_matrix}
	\frac{d}{d\tau}
	\begin{pmatrix}
		\eta \\
		\xi
	\end{pmatrix}
	=
	\begin{pmatrix}
		0 & A \\
		-\frac{1}{A} & -G
	\end{pmatrix}
	\begin{pmatrix}
		\eta \\
		\xi
	\end{pmatrix}.
\end{equation}
The Jacobian matrix in Eq.~(\ref{Jacobian_matrix}) has eigenvalues
\[\lambda=-\frac G2\pm i\sqrt{1-\frac{G^2}4}.\]
Since $G$ is positive and much smaller than $1$ (as shown in Appendix \ref{appendixB}, which estimates the experimental values of parameters such as $A$, $J$, and $G$ ), the term $ 1-G^2/4$  is positive. Therefore, the fixed point $(u^*,p^*)$ is a stable focus, and any trajectory in the plane will spiral towards this point. The perturbations $\eta(\tau)$ and $\xi(\tau)$ can be expressed as
\begin{subequations} \label{perturbations equations}
	\begin{align}
		&\eta(\tau) = C_1\exp(-\frac{G}{2}\tau)\cos(\sqrt{1-\frac{G^2}{4}}\tau+\Phi),\\
		&\xi(\tau)  = \frac1A\frac{du}{d\tau} = 		-C_1\exp(-\frac{G}{2}\tau)\cos(\sqrt{1-\frac{G^2}{4}}\tau+\Phi-\varphi),%\quad tan(\varphi)=\frac{\sqrt{1-\frac{G^2}{4}}}{\frac{G}{2}},
	\end{align}
\end{subequations} where $C_1$, $\Phi$ are undetermined coefficients related to initial values, and
\[
tan(\varphi)=\frac{\sqrt{1-\frac{G^2}{4}}}{\frac{G}{2}}.
\]
These results show that the perturbations in the ball's velocity and the gas pressure are both characterized by decaying oscillations with a frequency of $\sqrt{1-G^2/4}\sim1$. This indicates that in Koehler's experiment, whether the ball is below the small hole (in the half space ${y<0}$) or above it (in the half space $y\geq 0$), the changes in the ball's velocity and in the gas pressure remain closely approximated by harmonic motion, with a frequency nearly identical to the Rüchardt frequency $\omega$.

It is important to note that, so far, we can only state that the oscillation frequency of the system within each subspace is close to the Rüchardt frequency. However, we cannot arbitrarily say that the frequency of the overall motion, obtained by combining the trajectories from each subspace, is also close to the Rüchardt frequency. This is because the exact duration for which the system remains in each subspace is not known. For instance, let $T_0$ represent the Rüchardt period. If the system spends $T_0/2$ in one half-space and $T_0/3$ in another half-space, which may be possible, the total period of the motion will be $(T_0/2+T_0/3)$, not $T_0$, resulting in a frequency that deviates from the Rüchardt frequency. In the following section, we present a geometric analysis to demonstrate that, under appropriate experimental conditions, when the system alternates between the two half-spaces, the overall motion period approaches $T_0$, and the oscillation frequency remains close to the Rüchardt frequency.

\section{Periodic Solutions of Model Equations}
In this section, we estimate the time the system spends in each subspace using geometric considerations in the phase space and show that the overall period of motion approaches $T_0$. We show that the original three-dimensional (3D) PWLD system with a single separation plane $\{y = 0\}$ can be equivalently simplified to a two-dimensional (2D) PWLD system with a single separation line, which is more analytically tractable. Although this section is somewhat lengthy, it essentially describes how the system's trajectory revolves around two alternating focal points in a periodic manner.

\subsection{Difficulties in analyzing periodic solutions of the 3D PWLD system}
We first examine the difficulty of rigorously analyzing the periodic solutions of system~(\ref{matrix form 3D_equation}). Poincaré mapping\cite{strogatz2024nonlinear} is a powerful tool for studying periodic solutions in dynamical systems. However, for general PWLD systems, constructing a Poincaré map requires careful treatment of transitions across the switching surface within each subspace, which presents significant challenges. For system (\ref{matrix form 3D_equation}), we treat the separation plane $\{y = 0\}$ as the Poincaré section, which is a hypersurface transversally crossed by the system trajectories such that successive intersections define a discrete return map. In the present system, trajectories cross this plane with nonzero velocity in the $y$ direction, ensuring that the induced Poincaré map is well-defined. As discussed previously, when the system operates in either half-space, the velocity of the ball can be represented as
\begin{equation}\label{u_calculation}
  u=u^*+\eta(\tau)= u^*+C_1\exp(-\alpha\tau)\cos(\Omega\tau+\Phi),
\end{equation}
where $\alpha=G/2$, $\Omega=\sqrt{1-G^2/4}$. According to Eq. (\ref{dimensionless_dynamics}a), the coordinates of the ball can be obtained as
\begin{equation}\label{u_calculation}
\begin{aligned}
	y(\tau)&=y_{0}+\int_{0}^{\tau}u(t)dt\\&=y_{0}+u^*\tau+\frac{C_{1}}{\alpha^{2}+\Omega^{2}}\left\{-\alpha e^{-\alpha\tau}\cos(\Omega\tau+\Phi)+\Omega e^{-\alpha\tau}\sin(\Omega\tau+\Phi)+\alpha\cos(\Phi)-\Omega\sin(\Phi)\right\}.
	\end{aligned}
\end{equation}
By solving $y(\tau)=0$, in principle, the state of the system when switching between subspaces and its running time in each subspace can be obtained. However, the condition $y(\tau)=0$ is generally too complex to yield closed-form analytical solutions, which complicates the construction of Poincaré maps for analyzing periodic solutions. Although such maps can be constructed numerically, this typically requires case-by-case computation, limiting the possibility of obtaining general analytical insight.

\subsection{Simplified problem: 2D PWLD system with one separation line}
From the analysis above, it is evident that the system's trajectory in either half-space, when projected onto the $(u, p)$ plane, appears as a simple spiral around the focus. To avoid the issue of switching between subspaces, we reduce system~(\ref{matrix form 3D_equation}) to a 2D PWLD system:
\begin{equation} \label{matrix form 2D_equation}
	\dot{\mathbf{z}}=\begin{cases}M_+\mathbf{z}+\mathbf{e}_+&\mathrm{if}\quad p\leq0 \\M_-\mathbf{z}+\mathbf{e}_-&\mathrm{if}\quad p>0 \end{cases}
\end{equation}
with 
\[\mathbf{z}=\begin{pmatrix}u\\p\end{pmatrix},\quad
M_+=\begin{pmatrix}0&A\\-\frac1A&-G_+\end{pmatrix},\quad M_-=\begin{pmatrix}0&A\\-\frac1A&-G_-\end{pmatrix},\] and
\[ \mathbf{e}_+=\begin{pmatrix}0\\u^*_+\end{pmatrix},\quad
\mathbf{e}_-=\begin{pmatrix}0\\u^*_-\end{pmatrix},\]
where
\[
u^*_+ = A\left[ J - G_+(p_0 - 1) \right] < 0 , \qquad
u^*_- = A\left[ J - G_-(p_0 - 1) \right] > 0 .
\]
\begin{figure}[h!]
	\includegraphics[width = 0.75\textwidth]{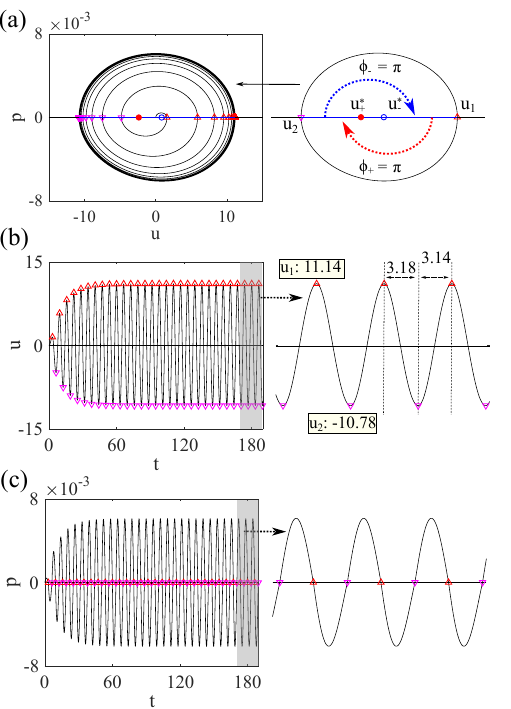}
	\caption{
		(Color online) (a) Phase portrait of the 2D system with separation line $\{p=0\}$,	
		showing the full dynamics including the transient regime (left), and the resulting periodic steady-state limit cycle with trajectory switching across the separation line (right). The numerical parameters are $A=1782$, $G_{+}=0.225$, $G_{-}=0.075$, and $J=1.0628\times10^{-3}$. The red solid and blue hollow circles denote the two fixed points, while the hollow upward and downward triangles indicate the intersections of the trajectory with the separation line. 
		(b)-(c) Oscillations of $u$ and $p$, respectively. The right-hand portions show enlarged views of the shaded regions.
	}
 
	\label{Phase portrait and oscillations of 2D with p=0}
\end{figure}

The periodic solutions of system (\ref{matrix form 2D_equation}) can be intuitively understood. Figure \ref{Phase portrait and oscillations of 2D with p=0} illustrates a typical periodic solution, with parameters \(A=1782\), \( G_{+} = 0.225 \), \( G_{-} = 0.075 \), and \( J = 1.0628 \times 10^{-3} \). In all numerical solutions presented in this paper, the values of $G_{\pm}$ and $A$ are kept fixed; these values are based on typical experimental estimates (see Appendix~\ref{appendixB}). Figure~\ref{Phase portrait and oscillations of 2D with p=0}(a) shows the phase portrait of the full dynamics including the transient regime (left), and the resulting periodic steady-state limit cycle, illustrating the trajectory switching between subspaces (right). In each half-plane, the trajectory is well-defined, spiraling clockwise around a focus. When the trajectory in $p>0$ intersects the separation line \( p = 0 \) at point \( (u_1, 0) \), it then enters the half-plane \( \{ p \leq 0 \} \), continuing its clockwise spiral while asymptotically approaching the focus \( (u_{+}^{*}, 0) \). It intersects the line \( p = 0 \) again at \( (u_2, 0) \). During this phase, the rotation radius, defined as the instantaneous Euclidean distance in phase space between the trajectory point $(u(t), p(t))$ and the fixed point $(u_{+}^{*}, 0)$, decays exponentially from $(u_1 - u_{+}^{*})$ to $(u_{+}^{*} - u_2)$. This quantity corresponds to the oscillation amplitude.

Subsequently, the system enters the half-plane \( \{ p > 0 \} \), switching its focus to the point \( (u_{-}^{*}, 0) \) and continuing its spiral towards this new focus. Since \( u_{-}^{*}<u_{+}^{*} \), the distance \( (u_{-}^{*} - u_2) \) is greater than \( (u_{+}^{*} - u_2) \). Thus, when the system switches subspaces, the rotation radius increases abruptly, compensating for the previous decay. Starting with \( (u_{-}^{*} - u_2) \) as the new rotation radius, the trajectory spirals clockwise around \( (u_{-}^{*}, 0) \) until it intersects the line \( p = 0 \) once more. For the system to maintain a periodic steady state, the trajectory must return to the point \( (u_1, 0) \), completing one full cycle of motion.

Remarkably, \( p = 0 \) is the line connecting the two foci. From a geometric perspective, it is evident that the trajectory in either half-plane rotates through an angle of \( \pi \), as shown in Fig. \ref{Phase portrait and oscillations of 2D with p=0}(a). Specifically, the system spends $T_\pm/2$  in each half-plane, where $T_\pm=2\pi/\Omega_\pm \sim 2\pi$, with $\Omega_\pm=\sqrt{1-G_\pm^2/4}\sim 1$ for $ G_{\pm}\ll1$. Therefore, the nondimensional period of the motion remains approximately $2\pi$, equal to the reduced Rüchardt period $T_0$. Having established the running time in each subspace, we now proceed to calculate the amplitude corresponding to the periodic solution as follows: 
\begin{equation}\label{u1_u2}
	  \begin{aligned}
	&u_{2}=u_{+}^{*}+(u_{+}^{*}-u_{1})e^{-\frac{G_{+}}{2}\cdot\frac{T_{+}}{2}}, \\ &u_{1}=u_{-}^{*}+(u_{-}^{*}-u_{2})e^{-\frac{G_{-}}{2}\cdot\frac{T_{-}}{2}}.
	\end{aligned}
\end{equation}
In the limit that \( G_{\pm}\ll1 \),
\begin{equation}\label{u1}
	u_{1}=\frac{u_{-}^{*}(1+e^{-\frac{G_{-}}{2}\cdot\frac{T_{-}}{2}})-u_{+}^{*}e^{-\frac{G_{-}}{2}\cdot\frac{T_{-}}{2}}(1+e^{-\frac{G_{+}}{2}\cdot\frac{T_{+}}{2}})}{1-e^{-(\frac{G_{+}}{2}\cdot\frac{T_{+}}{2}+\frac{G_{-}}{2}\cdot\frac{T_{-}}{2})}}\sim\frac{2(u_{-}^{*}-u_{+}^{*})}{\pi(G_{+}+G_{-})}.
\end{equation}

Substituting the parameters ($J,G_{\pm})$ used in Fig. \ref{Phase portrait and oscillations of 2D with p=0}, we calculate the following results: 

\begin{equation}\label{T/2}
\frac{T_{+}}{2}=3.18,\quad \frac{T_{-}}{2}=3.14, \quad T=6.32, 	
\end{equation}
and 
\begin{equation}\label{calculation results of u1 and u2}
	u_{1}=11.14, \quad u_{2}=-10.78.	
\end{equation}
These results agree exactly with the numerical simulation shown in Fig. \ref{Phase portrait and oscillations of 2D with p=0}(b).

According to Eq.(\ref{u1}) and Eqs.(\ref{u1_u2}), when $G_{\pm}\ll1$, the rotation radius of the trajectory can be approximated as $(u_{1}-u_{2})/2 \sim u_{1}-u_{+}^{*} >u_{1} \gg u_{-}^{*}-u_{+}^{*} $. Even if the separation line is not the one passing through the two foci (i.e., ${p=p_1\neq 0}$), the conclusion that the trajectory’s average rotation radius is significantly larger than the distance between the two foci remains valid. This indicates that, over one period, the trajectory can be approximated by a complete rotation around an effective center located between the two foci, corresponding to an angle of $2\pi$. Consequently, the system's period remains close to the Rüchardt period, although the deviation becomes slightly more pronounced as $p_1$ moves away from zero.

\begin{figure}[h!]
	\includegraphics[width = 0.75\textwidth]{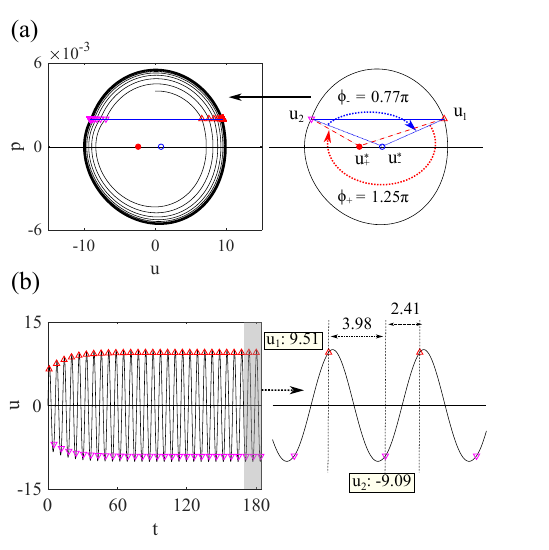}
	\caption{
		(Color online) (a) Phase portrait of the 2D system with separation line $\{p=0.002\}$, showing the full dynamics including the transient regime (left), and the resulting periodic steady-state limit cycle with trajectory switching across the separation line (right). All other parameters and graphical conventions are the same as in Fig.~\ref{Phase portrait and oscillations of 2D with p=0}. 
		(b) Oscillations of $u$. The right-hand portion shows an enlarged view of the shaded region.
	}

	\label{Phase portrait of 2D with y=1.2E-3}
\end{figure}

For instance, Figure \ref{Phase portrait of 2D with y=1.2E-3} illustrates the periodic solution when the separation line is given by $p=0.002$. Within a complete cycle, the system resides in the half-planes ${p \leq 0}$ and ${p > 0}$ for durations of $3.98$ and $2.41$, respectively, with corresponding rotation angles of $1.25~\pi$ and $0.77\pi$. The total rotation angle for the cycle is approximately $2.02~\pi$, demonstrating that the trajectory still maintains a near-complete $2\pi$ rotation despite the shifted separation line.

\subsection{The original 3D PWLD system with one separation plane y=0}
Returning to the original system described by Eqs. \ref{matrix form 3D_equation}, we now consider the case in 3D space with a separation plane $\{y=0\}$. For the most typical and simplest periodic solution, the trajectory crosses the plane $\{y=0\}$ once upwards and once downwards within one period, intersecting it at two distinct points, denoted as $A$ and $B$. On the projection plane $(u, p)$, let $A^{*}$ and $B^{*}$ represent the projections of points $A$ and $B$, respectively. At these two points, the projection of the periodic orbit transitions between subspaces. This correspondence allows the separation plane $\{y=0\}$ in the 3D system to be effectively reduced to a straight line connecting $A^{*}$ and $B^{*}$ in the 2D system. As a result, the periodic solutions of the original 3D system can be studied entirely within this simplified 2D framework. Since the periodic behavior of such 2D PWLD systems has been analyzed in the previous section, the periodic properties of the 3D system can be deduced accordingly. Specifically, a periodic solution that alternates between the half-spaces ${y\geq 0}$ and ${y<0}$ during a complete cycle retains a period that closely approximates the Rüchardt period.
\begin{figure}[h!]
	\includegraphics[width = 0.75\textwidth]{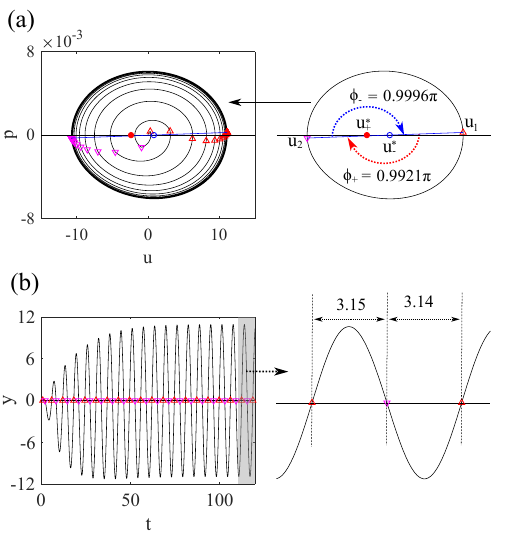}
    \caption{
	(Color online) (a) Projected phase portrait of the 3D system with separation plane $\{y=0\}$. The numerical parameters are $A=1782$, $G_{+}=0.225$, $G_{-}=0.075$, and $J=1.0628\times10^{-3}$. The hollow upward and downward triangles denote the projections onto the $(u,p)$ plane of the intersections where the trajectory crosses the separation plane $\{y=0\}$ upward and downward, respectively. As the trajectory approaches the steady-state periodic orbit, these two sequences of points converge to $u_1$ and $u_2$.
	(b) Oscillations of the displacement $y$, showing symmetry about the plane $\{y=0\}$. The right-hand portion shows an enlarged view of the shaded region. 
   }

	\label{Phase portrait and oscillations of 3D with symmetric y oscillation}
\end{figure}

\begin{figure}[h!]
	\includegraphics[width = 0.75\textwidth]{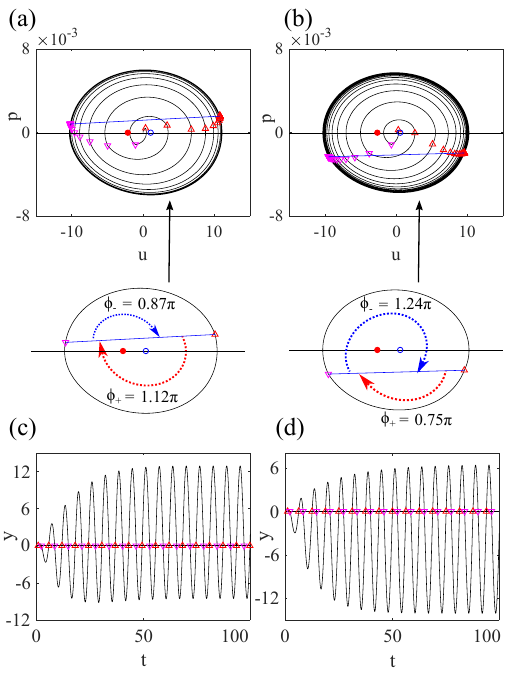}
    \caption{
	(Color online) Each column contains, from top to bottom, the projected phase portrait of the 3D system with separation plane $\{y=0\}$ and the oscillations of the displacement $y$. The parameters are $A=1782$, $G_{+}=0.225$ and $G_{-}=0.075$. For the left column, $J=1.2\times10^{-3}$, while for the right column $J=0.9\times10^{-3}$. For these parameter sets, the displacement $y$ exhibits asymmetric oscillations with respect to the plane $\{y=0\}$.
    }

	\label{Phase portrait and oscillations of 3D with asymmetric y oscillation}
\end{figure}
Figures~\ref{Phase portrait and oscillations of 3D with symmetric y oscillation} and~\ref{Phase portrait and oscillations of 3D with asymmetric y oscillation} present the numerical results of the 3D system (\ref{matrix form 3D_equation}) under different \( J \) (pumping). Figure~\ref{Phase portrait and oscillations of 3D with symmetric y oscillation} shows the case where the oscillation is symmetric about the plane $\{y=0\}$. When the ball passes through the small hole position, the pressure inside the container approaches the equilibrium pressure $P_0$, corresponding to the dimensionless reduced pressure $p\sim0$. As shown in Fig.~\ref{Phase portrait and oscillations of 3D with symmetric y oscillation}(a), the equivalent separation line in the projected $(u,p)$ plane nearly coincides with the line connecting the two foci, resembling the motion shown in Fig.~\ref{Phase portrait and oscillations of 2D with p=0}. Over a complete period, the trajectory rotates through angles of $ 0.9921~\pi$ and $ 0.9996~\pi $ in the two half-planes, respectively. The corresponding running time 3.15 and 3.14 are very close to $ \frac{T_{+}}{2}~(3.18) $ and $ \frac{T_{-}}{2}~(3.14)$, respectively, and the entire period $T=6.29$ is very close to the reduced Rüchardt period of $2\pi$.

Figure~\ref{Phase portrait and oscillations of 3D with asymmetric y oscillation} illustrates two cases of asymmetric oscillation of the displacement $y$ about the plane $\{y=0\}$. In these cases, the equivalent separation line deviates from the line connecting the two foci, resembling the motion shown in Fig.~\ref{Phase portrait of 2D with y=1.2E-3}. Over a complete period, the trajectory no longer rotates through angles close to $\pi$ in each half-plane. However, the total angular displacement still approaches $2\pi$.

In Koehler's experiment, it is essential to adjust the gas flow rate to ensure that the small ball oscillates symmetrically about the position of the hole. In the theoretical analysis presented in Koehler's paper \cite{Koehler1950}, it is assumed that the small ball spends equal amounts of time above and below the hole during each oscillation cycle. This assumption is necessary for carrying out the integration required for Fourier series analysis. The analysis in our paper clarifies the physical meaning of this requirement, showing that the symmetry about $y=0$ is not a strict requirement, but rather a limiting condition: the closer the motion is to being symmetric about $y=0$, the closer the oscillation period approaches the ideal Rüchardt period.

\section{Conclusion}

Koehler’s experiment provides a simple yet instructive realization of self-sustained nonlinear oscillations in a gas–mechanical system. Our analysis shows that, under conditions where the motion remains periodic and the small ball crosses the hole position once in each direction during a cycle, the resulting oscillation frequency is close to the classical Rüchardt frequency.

The measurement of $\gamma$ is typically introduced in lower-level undergraduate laboratory courses. The nonlinear dynamical concepts involved in the present work may therefore present a certain mathematical barrier for many students. For this reason, in practical teaching, even when discussing Koehler's experiment, it is useful to first adopt the physical framework underlying the Rüchardt experiment as the primary pedagogical model. After establishing this baseline understanding, the differences between the two experimental configurations can be highlighted. More advanced students can be encouraged to reproduce and extend the analysis presented in this work. This approach may help deepen their understanding of periodic motion and provide a concrete context for studying dynamical systems.

Finally, when multiple self-sustained oscillators are coupled, synchronization phenomena may emerge, as exemplified by the well-known synchronization of metronomes\cite{10.1119/1.1501118}. Introducing coupling between multiple Koehler apparatuses to demonstrate synchronization would be an interesting extension. In contrast to metronomes, whose operation is limited by a finite energy reservoir (the spring-driven mechanism), the oscillations in the Koehler system can, in principle, be sustained for an arbitrarily long time under continuous driving, making it a potentially robust platform for studying synchronization phenomena.

\section*{Acknowledgments}
This work was funded by the Research Project Supported by Shanxi Scholarship Council of China (Grant No.~2023-033) and by the Fundamental Research Program of Shanxi Province (Grant No.~202303021221071).

\section*{Author Declarations}
\textbf{Conflict of Interest}

The authors have no conflicts of interest to disclose.\\

\textbf{Author Contributions}

All authors contributed equally to this work.

\appendix  
\setcounter{figure}{0}
\renewcommand{\thefigure}{A\arabic{figure}}

\section{Details of our experimental setup}\label{appendixA}

\begin{figure}[h!]
	\includegraphics[width = 0.85\textwidth]{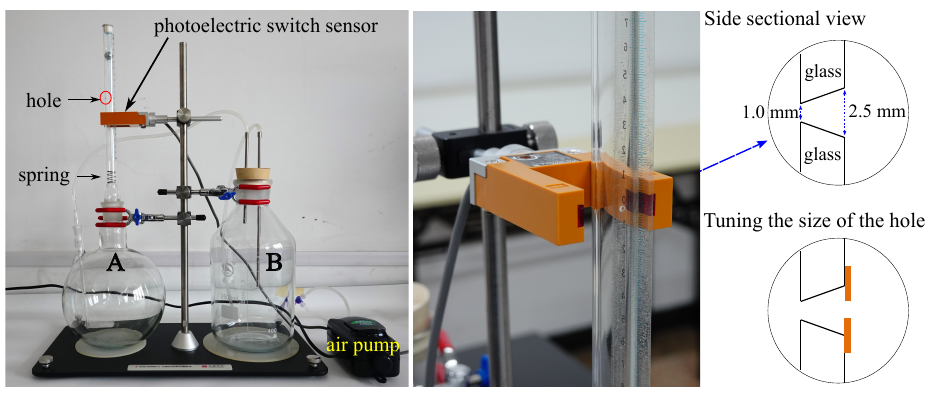}
	\caption{Details of our experimental setup.}
	\label{setup details}
\end{figure}

A key requirement for realizing the Koehler apparatus is achieving an appropriate match between the ball and the glass tube, with the difference in their diameters being only several tens of micrometers. Figure~\ref{setup details} shows the details of our experimental setup, which is based on a commercially available system~\cite{HXWest}. The ball is a bearing steel ball with a diameter of 14.00~mm and a mass of 11.40~g. The inner diameter of the glass tube is 14.05~mm. The tube and the fine gas inlet tube were connected to vessel A ($V_0=2.5\,\mathrm{L}$)  through ground-glass joints, providing airtight sealing and convenient assembly. The hole in the wall of the tube was drilled with a wide-angle drill, with an inner diameter of approximately 1.0 mm and an outer diameter of approximately 2.5 mm. As illustrated in the figure, the specific shape of the hole is not a necessary requirement but is merely intended to facilitate manufacturing. If adjustment of the hole size is required, a thin sheet with a smaller aperture can be used to block the original hole. The air pump employed is an aquarium air pump (model SB-748, Zhongshan SOBO Electric Appliance Co., Ltd., China), with a rated power of 8~W, a maximum air flow rate of 7~L/min, and a price of less than \$10. Vessel B serves as a gas reservoir, functioning to buffer and reduce pressure, thereby mitigating the effects of uneven air delivery from the air pump.

For constructing the experimental setup from individual components rather than using a pre-assembled commercial system, we recommend purchasing glass or acrylic tubes with a nominal inner diameter of 14~mm. Due to manufacturing tolerances, commercially available tubes often exhibit a slight positive deviation in inner diameter, which may be as large as approximately 14.10~mm in some cases.  Two practical approaches can be used to achieve an appropriate ball–tube matching condition. The first approach is to directly select steel balls with a diameter close to the expected range (e.g., 14.15~mm), which can be used as received. The second approach is to start from a slightly oversized steel ball (e.g., 14.20~mm) and gradually reduce its diameter by polishing or fine grinding until a suitable match with the tube is obtained. The adequacy of the matching can be assessed experimentally by observing whether the ball can be lifted by the air flow within the tube.

\section{Estimating the values of parameters in Eqs. (\ref{dimensionless_dynamics})}\label{appendixB}

\begin{figure}[h!]
	\includegraphics[width = 0.8\textwidth]{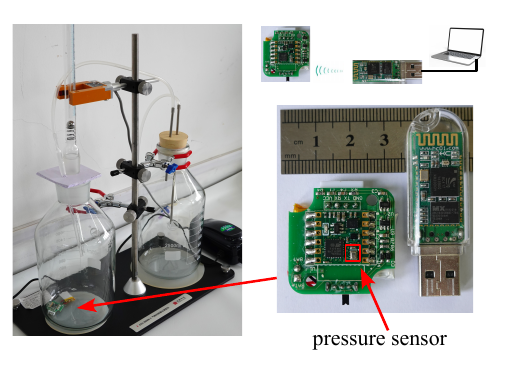}
	\caption{Measurement of the pressure variation inside the container using a wireless pressure sensor. Compared with the setup in Fig.~\ref{setup details}, a 2.5~L wide-mouth bottle was used as container A to accommodate the sensor. The container and glass tube were connected through an adapter plate, with the gaps sealed by hot-melt adhesive. The demonstration setup shown here was not finally sealed.}
	\label{P_measurment}
\end{figure}

The air pressure inside the vessel was measured using a miniature wireless Bluetooth multisensor module (WT901BC, WitMotion Shenzhen Co., Ltd., China)~\cite{WitMotion_WT901BC}, costing approximately \$30. Besides air pressure, the module can also measure the three-axis magnetic field, acceleration, and angular velocity. This module incorporates an SPL06-001 barometric pressure sensor (Goertek Microelectronics Inc., China), with a relative accuracy of $\pm 0.06$~hPa and a typical absolute accuracy of $\pm 1$~hPa. The module was placed directly inside the vessel, as illustrated in Fig.~\ref{P_measurment}. 

\begin{figure}[h!]
\includegraphics[width = .8\textwidth]{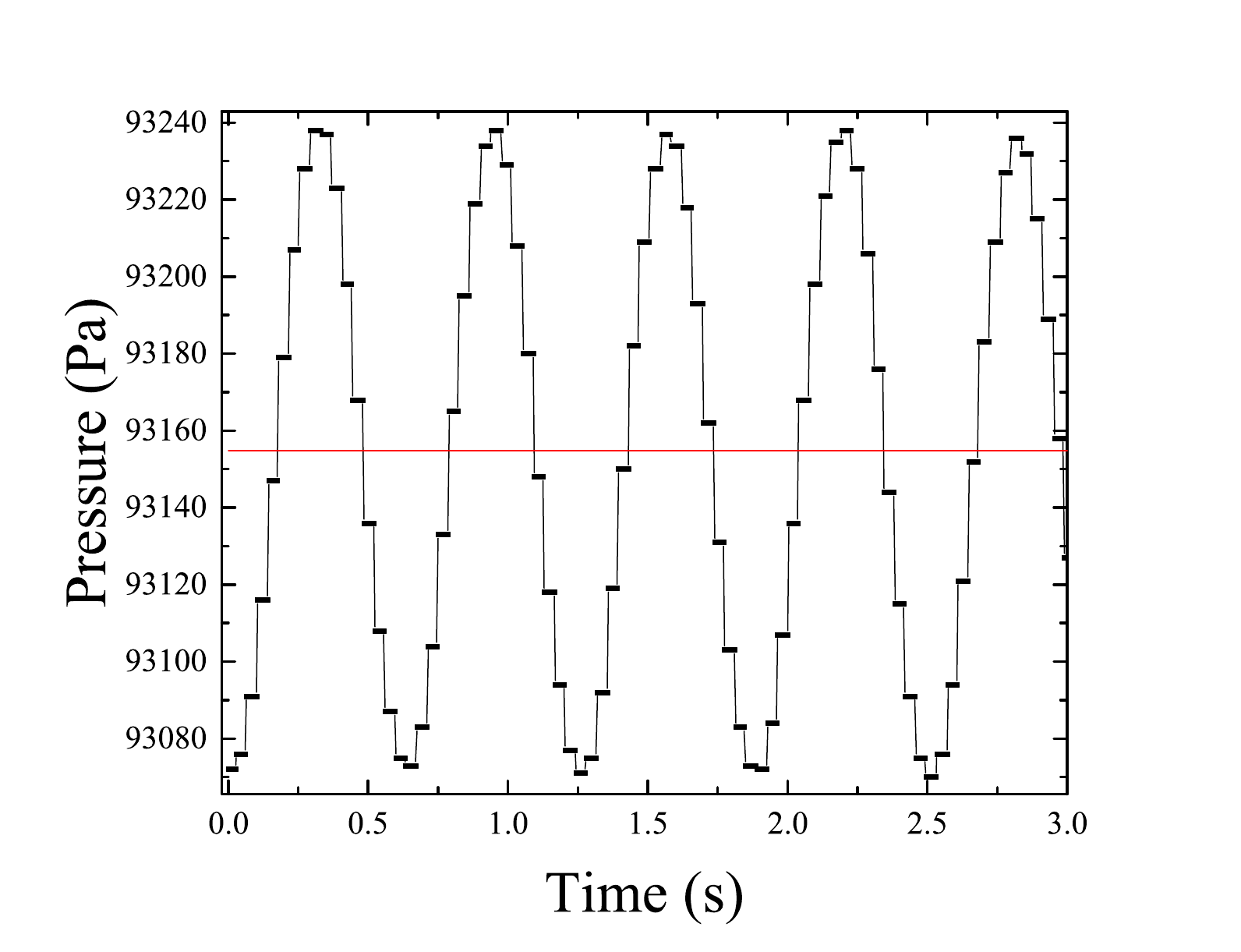}
\caption{Measured time evolution of the air pressure inside the container during self-sustained oscillations of the ball. The mean value of the pressure corresponds to the equilibrium pressure $P_0$ and is shown with a horizontal line.}
\label{P-t}
\end{figure}

Figure~\ref{P-t} shows the temporal variation of the air pressure inside the container during the self-sustained oscillations of the ball. The atmospheric pressure was $P_{\mathrm{atm}} = 92400~\mathrm{Pa}$. The equilibrium pressure was approximately $P_{0} \sim 93155~\mathrm{Pa}$, the oscillation period was $T \sim 0.6~\mathrm{s}$, and the corresponding angular frequency was $\omega \sim 10~\mathrm{rad/s}$. One can then determine the dimensionless parameters $p_0=P_0/P_{\mathrm{atm}}=1.008$, and $A=(\pi RP_{\mathrm{atm}})/(m\omega^{2})=1782$. The pressure variations within half a period is approximately $170\,\mathrm{Pa}$, so $\partial P/\partial t=jP_{\mathrm{atm}}-g(x)(P-P_{\mathrm{atm}})$ is of the order of magnitude of $567~\mathrm{Pa\,s^{-1}}$. Considering that the contributions from gas supply and leakage are comparable in magnitude, it is appropriate to assign gas leakage parameters
\[ e(P_0-P_{\mathrm{atm}})\sim567~\mathrm{Pa\,s^{-1}}, \]
\[ (e+f)(P_0-P_{\mathrm{atm}})\sim 3\times 567~\mathrm{Pa\,s^{-1}},\]
where the numerical factor $3$ is introduced as a representative choice for numerical estimation, without implying a specific relation between $e$ and $f$.
Then we obtain
\[e\sim0.75\, \mathrm{s}^{-1},\,f\sim1.5\,\mathrm{s}^{-1},\]
and the dimensionless $G(y)$ is \[G(y)=\begin{cases}G_+ = (e+f)/\omega= 0.225,\quad&y\geq 0,\\G_- = e/\omega= 0.075,\quad&y<0.\end{cases}\]
Then, from Eq.~\eqref{fixed_point}, the fixed point $(u^*, p^*)$ satisfies
\[
u^* = A\left[J - G_{\pm}(p_0 - 1)\right].
\]
Let $J$ (the gas supply parameter) be taken as the control parameter. For self-sustained oscillations to occur, $J$ must satisfy
\[
\begin{aligned}
	u^*_+ &= A\left[ J - G_+(p_0 - 1)\right] < 0, \\
	u^*_- &= A\left[ J - G_-(p_0 - 1)\right] > 0.
\end{aligned}
\]

\FloatBarrier

\bibliographystyle{unsrt}
\bibliography{reference.bib}

\end{document}